\documentclass[aps,prd,onecolumn,groupedaddress,showpacs,nofootinbib,amssymb]{revtex4}
\usepackage[dvips]{graphicx}
\usepackage{amssymb}
\usepackage{amsmath}
\usepackage{graphicx}
\usepackage{amsfonts}
\usepackage{bm}
\usepackage{latexsym,float}
\usepackage{epsfig}
\usepackage{epstopdf}
\usepackage{amssymb}
\usepackage{verbatim}
\usepackage{multirow}
\usepackage{color}

\begin{document}

\title{Phase Space Analysis of the Accelerating Multi-fluid Universe }

\author{S.D.~Odintsov$^{1,2}$}
\email{odintsov@ieec.uab.es} \affiliation{$^{1)}$Institut de
Ciencies de lEspai (IEEC-CSIC),  Carrer de Can Magrans, s/n, 08193
Barcelona, Spain \\ $^{2)}$ICREA, Passeig LluAs Companys, 23,
    08010 Barcelona, Spain}

\author{V.K.~Oikonomou$^{3,4}$}
\email{v.k.oikonomou1979@gmail.com} \affiliation{$^{3)}$Laboratory
for Theoretical Cosmology, Tomsk State University of Control Systems
and Radioelectronics (TUSUR), 634050 Tomsk, Russia\\ $^{4)}$Tomsk
State Pedagogical University, 634061 Tomsk, Russia}

\author{Petr V. Tretyakov$^{5,6}$}
\email{tpv@theor.jinr.ru} \affiliation{$^{5)}$Joint Institute for
Nuclear Research, Joliot-Curie 6, 141980 Dubna, Moscow region,
Russia\\ $^{6)}$Institute of Physics, Kazan Federal University,
Kremlevskaya street 18, 420008 Kazan, Russia}

\begin{abstract}
We study in detail the phase space of a Friedmann-Robertson-Walker
Universe filled with various cosmological fluids which may or may
not interact. We use various expressions for the equation of state,
and we analyze the physical significance of the resulting fixed
points. In addition we discuss the effects of the stability or an
instability of some fixed points. Moreover we study an interesting
phenomenological scenario for which there is an oscillating
interaction between the dark energy and dark matter fluid. As we
demonstrate, in the context of the model we use, at early times the
interaction is negligible and it starts to grow as the cosmic time
approaches the late-time era. Also the cosmological dynamical system
is split into two distinct dynamical systems which have two distinct
de Sitter fixed points, with the early-time de Sitter point being
unstable. This framework gives an explicit example of the
unification of the early-time with late-time acceleration. Finally,
we discuss in some detail the physical interpretation of the various
models we present in this work.
\end{abstract}

\pacs{04.50.Kd, 95.36.+x, 98.80.-k, 98.80.Cq, 11.25.-w }

\maketitle

\section{Introduction}

In modern theoretical cosmology, the most striking event was the
observation of the late-time acceleration \cite{riess} that our
Universe undergoes at present time. Admittedly this observation has
utterly changed the way of thinking of modern cosmologists, since
this late-time acceleration is a feature of our Universe that was
never thought it would actually occur. Consequently, the focus for
the last nearly 20 years is to model in a successful way this
late-time acceleration and also to harbor the late and early-time
acceleration era in a unified theoretical framework. Towards this
unified description, many proposals, especially those suggesting to
modify the gravitational sector, have been introduced ever since,
see the reviews \cite{reviews1,reviews2,reviews3,reviews4} for
details. From the first moment that the late-time acceleration has
been observed, it was realized that no perfect matter fluid known at
that time was able to realize the late-time acceleration era, and
therefore the need for alternative generalized cosmological fluids
was compelling. By using generalized cosmological fluids, both the
late and early-time acceleration era can be realized, and up to date
there are many theoretical proposals that use generalized fluids,
for example in Refs.
\cite{inhomogen1,nojodineos1,inhomogen2,brevik1,brevik2,fluid1,fluid2}
imperfect fluids are used in order to describe the cosmological
evolution of our Universe, and in some cases certain particular
examples are used, called viscous fluids are (see
\cite{Brevik:2017msy} for reviews). It is notable that the imperfect
fluids may describe even phantom evolution of our Universe, without
using phantom scalar fields, which violate the energy conditions,
see Refs. \cite{inhomogen1,inhomogen2} for details. Furthermore,
other cosmological evolutions like bouncing cosmology, in the
context of both classical and loop quantum cosmology imperfect
fluids were studied in \cite{oikonomouimperfect}, and also singular
cosmology can be realized by imperfect fluids
\cite{oikonomoufluid,Brevik:2016kuy}. Furthermore, several models
which take into account bulk viscosity were discussed in Refs.
\cite{ALUL,CLLPS,LS,VWM,VS,VCFCB,VSFZ,V,IS,M,CLP,CCL,LOS}, and also
an important class of models which assume an interaction between
dark matter and dark energy fluids, can be found in Refs.
\cite{ALUL,ZP,CCMUL}.

In this paper we shall perform a detailed phase-space analysis of a
Friedmann-Robertson-Walker Universe, filled with different fluid
components, which may or may not interact between them. The
dynamical evolution of such kind of model is described by the
Friedmann equations,
\begin{equation}
\frac{3}{\kappa^2}H^2=\sum\rho_i,
\label{0.1}
\end{equation}
\begin{equation}
\frac{-1}{\kappa^2}(2\dot H + 3H^2)=\sum p_i,
\label{0.1.1}
\end{equation}
or equivalently,
\begin{equation}
\frac{-2}{\kappa^2}\dot H = \sum(\rho_i+ p_i),
\label{0.1.2}
\end{equation}
and also the energy conservation equations hold true,
\begin{equation}
\dot\rho_i=-3H(\rho_i+p_i). \label{0.2}
\end{equation}
In the above, $\kappa^2=8\pi G$ and also the equation of state (EoS)
$p_i=p_i(\rho_i)$ may be highly non-trivial for some fluid
components. We shall appropriately choose the variables in order to
capture the phase space dynamics in the most optimal way, and we
shall analyze the structure of the phase space by providing an
analytic treatment of the cosmological dynamical equations. Seeing
the cosmological equations as a dynamical system is a particularly
appealing way to investigate the phase space in many cosmological
contexts, see for example
\cite{Coley:1999uh,GarciaSalcedo:2012dn,Boehmer:2008av,Boehmer:2014vea,Haba:2016swv,Bolotin:2013jpa},
but also in modified gravity too, see for example
\cite{Abdelwahab:2007jp,Carloni:2007eu,Carloni:2004kp,Leon:2014yua,Xu:2012jf,Carloni:2017ucm,Leon:2010pu,Carloni:2007br,Khurshudyan:2016qox,Hohmann:2017jao}.
In most cases, the choice of the dynamical system variables plays a
crucial choice, and in some cases the resulting cosmological
dynamical system may be rendered autonomous \cite{Odintsov:2015wwp}.
Also it is possible to choose dimensionless variables, see for
example Refs. \cite{Carloni:2007eu,Carloni:2004kp} for an $F(R)$
gravity cosmological dynamical system, and also see Ref. \cite{STT},
for a cosmological theory with higher derivatives of the scalar
curvature. The dynamical systems approach for cosmological systems
has many attributes, with the most important being the fact that the
fixed points of the dynamical system actually provide new insights
with regards to the behavior of the attractor solutions and also
reveals the stability structure of the dynamical system near the
attractors. It is conceivable that the choice of the variables plays
an important role, as we also demonstrate by this work.

This paper is organized as follows: In section \ref{sec:1} we
present some well known features of dynamical systems approach in
cosmological context with generalized fluids, in section \ref{sec:2}
we discuss in brief how interactions between dark matter and dark
energy may be introduced and by using appropriately chosen variables
we present how the dynamical systems analysis can be performed in
this case. In section \ref{sec:3} we generalize the formalism we
developed in the previous sections and by using dimensionless
variables, we investigate the physical consequences of having
various equations of state for the fluid components of the
cosmological system. In section \ref{sec:4} by using appropriately
chosen dimensionless variables, we investigate in detail how the
interaction of dark matter and dark energy fluids may affects the
phase space structure. We study the stability and behavior of the
fixed points of the dynamical system and we also discuss how the
early and late-time acceleration eras are affected by the various
functional forms of the interaction coupling between dark matter and
dark energy. Finally the conclusions follow in the end of the paper.

\section{Standard Approach on Dynamical Systems and Cosmological Dynamics}\label{sec:1}

In this section we present the simplest case of the dynamical
systems approach in cosmological dynamics. We consider the simplest
case in which the Universe is filled with radiation $\rho_r$ and a
perfect fluid with a non-trivial EoS of the form
$p=-\rho+f(\rho)+G(H)$ \cite{NO1}. Such non-trivial EoS can be
considered as some sort of viscous fluid or generalized EoS fluid,
see \cite{Bamba:2012cp} for a review on this topic. In addition,
such an EoS can be considered as an effective fluid presentation of
some modified gravity theory \cite{reviews2,reviews3}. In the case
at hand, the full dynamical system takes the following form,
\begin{eqnarray}
\frac{3}{\kappa^2}H^2&&=\rho_r+\rho=\rho_{tot},\label{1.1}\\
\dot\rho_r&&=-4H\rho_r,\label{1.2}\\
\dot\rho&&=-3H[f(\rho)+G(H)].\label{1.3}
\end{eqnarray}
The appearance of the term $G(H)$ might seem unconventional, from a
thermodynamic point of view, and we need to briefly describe the
motivation for using such a term. This term encompasses the viscous
part of the cosmological fluid, so it mainly quantifies the
viscosity of the fluid. In the Universe, nd especially in the very
early stages of it's evolution, the effects of a viscous
cosmological component are most likely expected to occur during the
neutrino decoupling process, which occurs at the end of the lepton
era \cite{Misner:1967uu}. Hence, viscosity is encompassed in the
very own fabric of the Universe. In addition, a strong motivation
for using viscous fluid components comes from the fact that the
perfect fluid approach among cosmologists-hydrodynamicists is just
an ideal approach, and does not describe the real world. Finally,
due to the fact that early and late-time acceleration may be
described by an unknown form of a cosmological fluid, it is natural
to assume that the fluid has the most general form, which means that
a viscous component is needed\footnote{Note that terms of the form
$G(H,\dot{H})$ in the cosmological relation between the effective
pressure and the energy density, naturally occur in modified gravity
thermodynamics \cite{Bamba:2016aoo}}.

Having described the motivation for the use of viscous fluid
components, we can rewrite the above cosmological equations in terms
of dimensionless variables. In the case at hand, there is only one
independent variable due to the constraint equation (\ref{1.1}). By
using the $e$-foldings number\footnote{In this case time derivatives
for some variable $X$
  transform as $X'=\frac{dX}{dN}=\frac{\dot X}{H}$} $N=\ln a$ and by introducing a new dimensionless
  variable defined as,
$$x=\frac{\kappa^2}{3H^2}\rho=\frac{\rho}{\rho_{tot}},$$
we obtain the next dynamical equation,
\begin{equation}
x'=\frac{3}{\rho_{tot}}\left [ f(\rho) + G(H) \right](x-1) -4x^2 +4x,
\label{1.4}
\end{equation}
where the variables $\rho$, $H$ and $\rho_{tot}$ must be expressed
in terms of $x$ depending on the choices of the functions $f$ and
$G$.

Let us here discuss the simplest choice of EoS, which is $p=w_0\rho
+w_1H^2$, which implies that $f(\rho)=\rho(1+w_0)$ and
$G(H)=w_1H^2$. It is easy to see that in this case, Eq. (\ref{1.4})
takes the following form,
\begin{equation}
x'=\left [ 3(1+w_0)x +\kappa^2w_1 \right](x-1) -4x^2 +4x\equiv m(x),
\label{1.5}
\end{equation}
where the ``prime'' denotes differentiation with respect to the
$e$-foldings number. The equation that determines the stationary
points for the above dynamical equation, is quadratic with
discriminant $\mathfrak{D}=\left(\kappa^2w_1 -4 + 3(1+w_0)
\right)^2$, so the existing solutions are always real. For the case
$\mathfrak{D}=0$ there is the only one solution, which is,
$x=\frac{1}{2}-\frac{\kappa^2w_1}{2(3(1+w_0)-4)}$. For the case
$\mathfrak{D}>0$ there are two solutions, which are:
\begin{eqnarray}
x_1&&=1,\label{1.6}\\
x_2&&=\frac{-\kappa^2w_1}{3(1+w_0)-4},\label{1.7}
\end{eqnarray}
and this case is the most interesting, since it provides us plenty
dynamical solutions. Now note that the physical values that the
variable $x$ can take, are $0\leqslant x\leqslant 1$, but for $x_2$
we have $0<x_2<1$, which give us the following restrictions for the
free parameters:
\begin{eqnarray}
&&w_1<0,\,\,w_0>-\frac{1}{4},\,\,-\kappa^2w_1<-4+3(1+w_0),\\
&&w_1>0,\,\,w_0<-\frac{1}{4},\,\,-\kappa^2w_1>-4+3(1+w_0).
\end{eqnarray}
Let us study the stability conditions of the existing stationary
points. It is clear that for every point $x_i$ there is only one
eigenvalue $\mu_i=m'(x=x_i)$, and for the fixed points at hand, we
have,
\begin{eqnarray}
\mu_1(x_1)&&=-4+3(1+w_0)+\kappa^2w_1,\\
\mu_2(x_2)&&=4-3(1+w_0)-\kappa^2w_1=-\mu_1 .
\end{eqnarray}
Therefore, we find that for any values of parameters $w_0$, $w_1$
one of the fixed points is stable and the other one is unstable.
Note that for the case $\mathfrak{D}=0$ we have $\mu=0$ and
therefore it is compelling to investigate the corresponding center
manifold, but fortunately this is not interesting case from a
physical point of view. The physical significance of the stationary
point $x_1$ is that it corresponds to a Universe with
$\rho_{tot}=\rho$ and $\rho_r=0$. With regard to the fixed point
$x_2$, it corresponds to a Universe with some fixed relation between
$\rho$ and $\rho_r$.

\section{Models with Dark Matter Interacting to Dark Energy}\label{sec:2}

\begin{figure}[h!]
  \includegraphics[width=10cm]{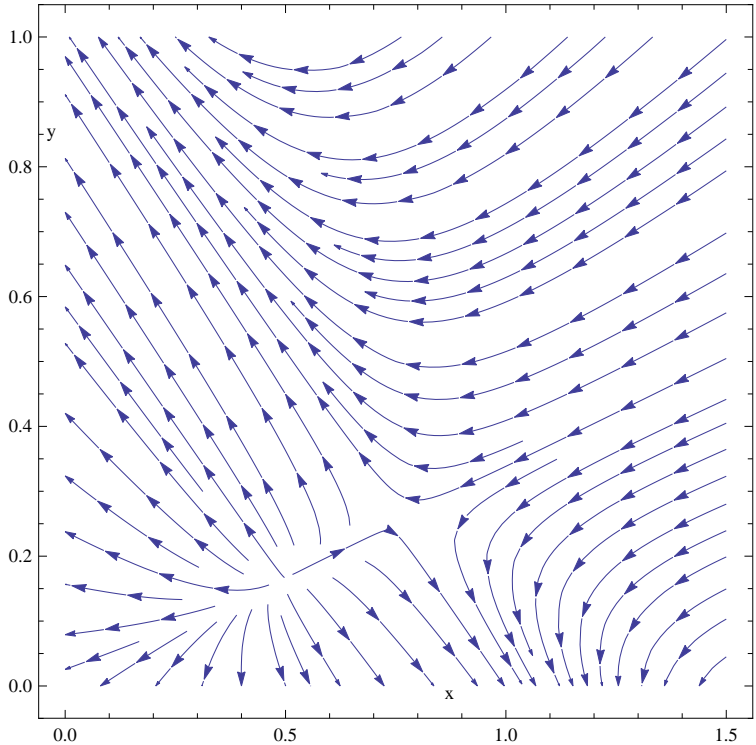}
  \caption{Phase portrait for the case $w_0=0$, $q=-\frac{1}{2}$, $w_1\kappa^2=1$, $\zeta_0=\frac{1}{9}$}
  \label{fig:fig0}
\end{figure}

Let us discuss at this point some models which describe interactions
of dark matter with dark energy. The dynamical system in this case
can be written in the following form,
\begin{eqnarray}
\frac{3}{\kappa^2}H^2&&=\rho_r+\rho+\rho_{dm}=\rho_{tot},\label{2.1}\\
\rho_r'&&=-4\rho_r,\label{2.2}\\
\rho'&&=-3[f(\rho)+G(H)]-\frac{Q}{H},\label{2.3}\\
\rho'_{dm}&&=-3\rho_{dm}+\frac{Q}{H}-3(-3H\zeta),\label{2.4}
\end{eqnarray}
where the prime denotes as previously differentiation with respect
to the $e$-foldings number $N$, and $Q$ quantifies the interaction
between dark energy and dark matter, and the term $-3H\zeta$
corresponds to the bulk viscous pressure of the dark matter fluid.

We will assume that the bulk viscous coefficient $\zeta$ has the
following form \cite{ALUL,Brevik:2005bj,Brevik:2007ig},
\begin{equation}
\zeta=\frac{\zeta_0}{\sqrt{3\kappa^2}}\rho_{tot}^{\frac{1}{2}}=\frac{1}{\kappa^2}H\zeta_0.
\label{2.5}
\end{equation}
The motivation for using this kind of ansatz, comes from
astronomical estimates on the relation between $\zeta$ and
$\rho_{tot}$, which was studied in Ref. \cite{Normann:2016jns},
where a general form of the $\zeta-\rho_{tot}$ relation was assumed,
and it was of the form $\zeta\sim \rho^{\lambda}_{tot}$. The choice
$\zeta\sim \rho^{1/2}_{tot} $ we used in Eq. (\ref{2.5}) yields a
good fit between astronomical constraints and the fluid approach.

In this case we introduce the following set of dimensionless
variables,
\begin{eqnarray}
x&&=\frac{\kappa^2}{3H^2}\rho=\frac{\rho}{\rho_{tot}},\label{2.6}\\
y&&=\frac{\kappa^2}{3H^2}\rho_{dm}=\frac{\rho_{dm}}{\rho_{tot}},\label{2.7}\\
q&&=\frac{\kappa^2}{3H^3}Q=\frac{Q}{H\rho_{tot}},\label{2.8}
\end{eqnarray}
where only $x$ and $y$ are independent dynamical variables. By using
the set of dimensionless variables, the dynamical equations
(\ref{2.1})-(\ref{2.4}) take the form\footnote{Note that the
constraint (\ref{2.1}) and the equation for $\rho_r$ are already
took into account in this system.}:
\begin{eqnarray}
x'&&=-4x^2+4x-yx-3x\zeta_0-q+\frac{3(f+G)}{\rho_{tot}}(x-1),\label{2.9}\\
y'&&=-y^2-4xy+y(1-3\zeta_0)+3\zeta_0+q+\frac{3(f+G)}{\rho_{tot}}y,\label{2.10}
\end{eqnarray}
and by using the EoS we used in the previous section, namely,
$f(\rho)=\rho(1+w_0)$, $G(H)=w_1H^2$ we obtain,
\begin{eqnarray}
x'&&=x^2(3w_0-1)+x(w_1\kappa^2-3\zeta_0-3w_0+1)-xy-(q+w_1\kappa^2),\label{2.11}\\
y'&&=-y^2+xy(3w_0-1)+y(w_1\kappa^2+1-3\zeta_0)+3\zeta_0+q.\label{2.12}
\end{eqnarray}
We can easily find the stationary points for the above dynamical
system, by multiplying (\ref{2.9}) by $y$, (\ref{2.10}) by $(1-x)$
and by combining the resulting equations we find,
\begin{equation}
(1-x_0-y_0)(y+3\zeta_0+q)=0.
\label{2.11.0}
\end{equation}

The above equation indicates that there exist at most three
stationary points, and it is clear that even in the most general
case, the stationary points may be found analytically, and these are
equal to,
\begin{equation}
x_0=\frac{q+w_1\kappa^2}{1-3w_0},\,\,
y_0=-3\zeta_0-q,
\label{2.11.0.0}
\end{equation}
\begin{equation}
x_0=1-y_0,\,\,
y_0=\frac{1}{6w_0}\left [ 3w_0+w_1\kappa^2-3\zeta_0 \pm\sqrt{(3w_0+w_1\kappa^2-3\zeta_0)^2+12w_0q+36w_0\zeta_0} \right ].
\label{2.11.0.1}
\end{equation}
Since there are a many free parameters, we investigate some
particular cases, which are interesting from a physical point of
view. We start off with the case $w=0$, in which case the first
fixed point takes the form,
\begin{equation}
y_0=-3\zeta_0-q,\,\,\,x_0=q+w_1\kappa^2,
\label{2.12.0}
\end{equation}
and the corresponding eigenvalues are,
\begin{eqnarray}
\mu_1&&=1,\label{2.13.0}\\
\mu_2&&=1-w_1\kappa^2+3\zeta_0,\label{2.14.0}
\end{eqnarray}
Accordingly, the second fixed point is,
\begin{equation}
y_0=\frac{3\zeta_0+q}{3\zeta_0-w_1\kappa^2},\,\,\,x_0=1-y_0,
\label{2.15}
\end{equation}
and the corresponding eigenvalues are,
\begin{eqnarray}
\mu_1&&=w_1\kappa^2-3\zeta_0,\label{2.16}\\
\mu_2&&=w_1\kappa^2-3\zeta_0-1,\label{2.17}
\end{eqnarray}
so the first point always unstable, whereas the second one may be
stable, depending on the choice of the free parameters. However, if
the first point lies in the physically allowed region, the second
fixed point is rendered always unstable. Typical phase portrait with
the two fixed points in the physical region, are presented in
Fig.\ref{fig:fig0}.

Now let us consider another physically interesting case, for which
$q=-3\zeta_0$. In this case, there exist three fixed points which
are,
\begin{eqnarray}
y_0&&=0,\,\,x_0=\frac{3\zeta_0-w_1\kappa^2}{3w_0},\,\,\mu_{1}=w_1\kappa^2-3\zeta_0+1+x_0(3w_0-1),\,\,\mu_{2}=2\mu_1-w_1\kappa^2+3\zeta_0-3w_0-1,\label{2.15.0}\\
y_0&&=0,\,\,x_0=1,\,\,\mu_{1}=w_1\kappa^2-3\zeta_0+3w_0,\,\,\mu_{2}=w_1\kappa^2-3\zeta_0+3w_0-1,\label{2.15.1}\\
y_0&&=\frac{3w_0+w_1\kappa^2-3\zeta_0}{3w_0},\,\,x_0=1-y_0,\,\,\mu_{1}=-1,\,\,\mu_{2}=  -3w_0-w_1\kappa^2+3\zeta_0.\label{2.16}
\end{eqnarray}
A particularly interesting subcase of the above is if we further
choose, $3\zeta_0=w_1\kappa^2$, then the fixed points
become,
\begin{eqnarray}
x_0&&=0,\,\,y_0=1,\,\,\mu_{1,2}=-1,\,\,-3w_0,\label{2.17}\\
x_0&&=1,\,\,y_0=0,\,\,\mu_{1,2}=3w_0,\,\,3w_0-1,\label{2.18}\\
x_0&&=0,\,\,y_0=0,\,\,\mu_{1,2}=1,\,\,1-3w_0.\label{2.19}
\end{eqnarray}
We can see that two fixed points are always unstable, whereas one of
the three, the first or the second one depending on the sign of the
parameter $w_0$, is stable.

\begin{figure}[h!]
  \includegraphics[width=10cm]{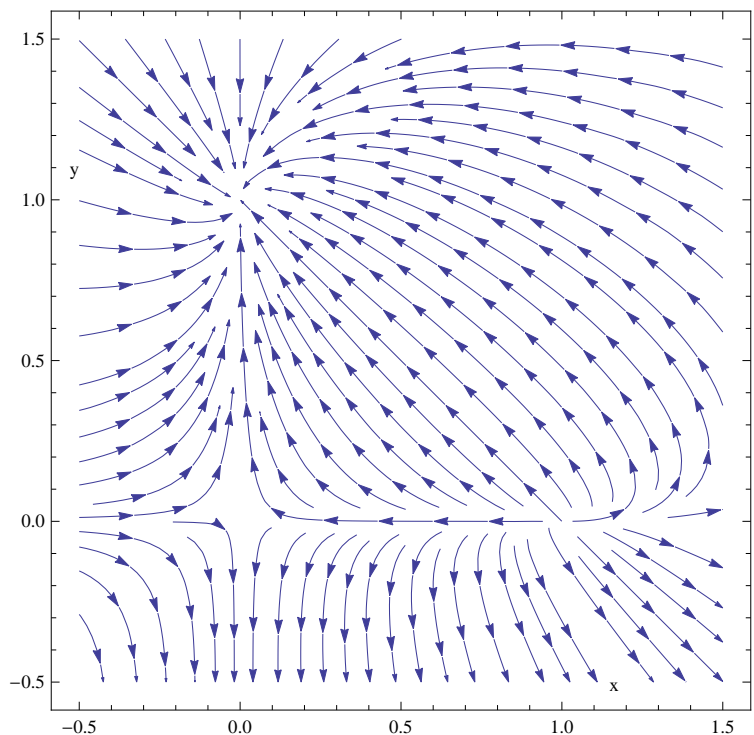}
  \caption{Phase portrait for the case $w_0=0.5$, $q=-3\zeta_0$, $q=-w_1\kappa^2$.}
  \label{fig:fig1}
\end{figure}

\begin{figure}[h!]
  \includegraphics[width=10cm]{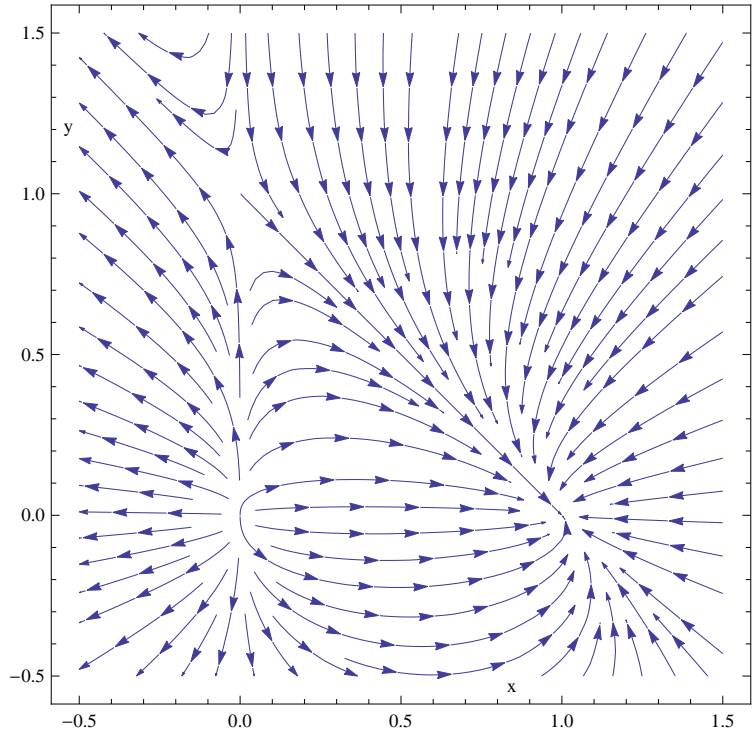}
  \caption{Phase portrait for the case $w_0=-0.5$, $q=-3\zeta_0$, $q=-w_1\kappa^2$}
  \label{fig:fig2}
\end{figure}

\begin{figure}[h!]
  \includegraphics[width=10cm]{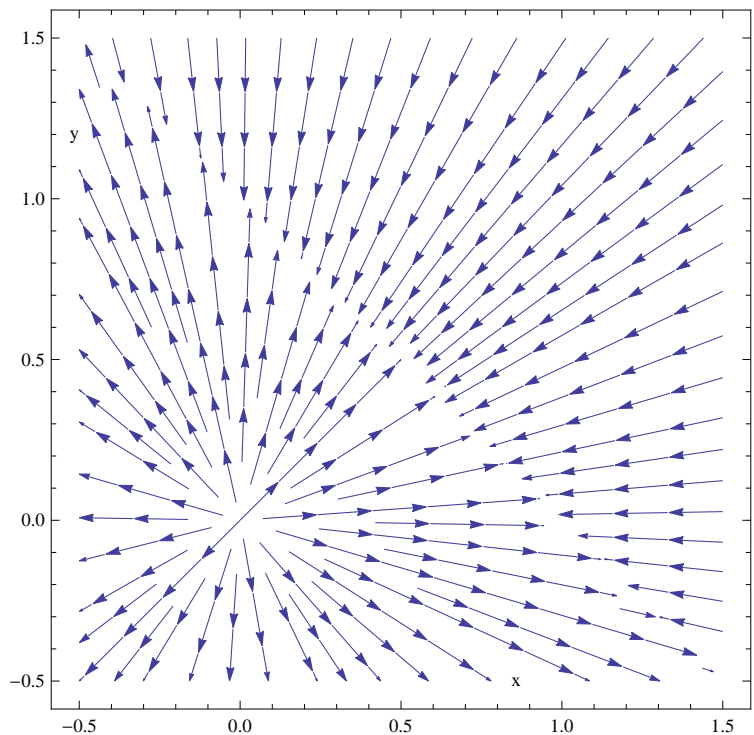}
  \caption{Phase portrait for the case $w_0=0$, $q=-3\zeta_0$, $q=-w_1\kappa^2$}
  \label{fig:fig3}
\end{figure}

In Fig. \ref{fig:fig1} we plotted the phase portrait for the case
$w_0=0.5$, while in Fig. \ref{fig:fig2} we plotted the phase
portrait for $w_0=-0.5$, and finally in Fig. \ref{fig:fig3} we can
see the phase portrait for $w_0=0$. As it can be seen in all
figures, two of the three fixed points are unstable and one of the
three is stable. Furthermore, it can be seen that by using Eq.
(\ref{2.11.0.1}), it is possible to find the case (by appropriately
choosing the parameters) for which there will be some stationary
point with fixed relation $x/y\neq0,1$. This task may be solved
numerically but we refrain from going into details on this.

\section{Generalized Form of Cosmological Fluids}\label{sec:3}

In this section we extend the cases we presented in section
\ref{sec:1} to include generalized form of the EoS. We consider the
simplest scenario for which the Universe is filled with $\rho_r$ and
a perfect fluid with non-trivial EoS $p=-\rho+f(\rho)+G(H)$
\cite{NO1}. In this case, the dynamical system takes the following
form,
\begin{eqnarray}
\frac{3}{\kappa^2}H^2&&=\rho_r+\rho=\rho_{tot},\label{3.1}\\
\frac{-2}{\kappa^2}\dot H &&= \frac{4}{3}\rho_r + [f(\rho)+G(H)],\label{3.1.1}\\
\dot\rho_r&&=-4H\rho_r,\label{3.2}\\
\dot\rho&&=-3H[f(\rho)+G(H)].\label{3.3}
\end{eqnarray}
By using the $e$-foldings number as independent variable and also by
introducing the dimensionless variables,
$$x=\frac{\kappa^2}{3H^2}\rho=\frac{\rho}{\rho_{tot}},$$
$$z=\frac{1}{\kappa^2H^2}=\frac{3}{\rho_{tot}\kappa^4}$$
the dynamical system can be cast in the following form,
\begin{eqnarray}
x'&&=\kappa^4z\left [ f(\rho) + G(H) \right](x-1) -4x^2 +4x,\label{3.4.1}\\
z'&&=4z(1-x)+z^2\kappa^4\left [ f(\rho) + G(H) \right],\label{3.4.2}
\end{eqnarray}
where according to the definition, $\rho=\frac{3x}{\kappa^4z}$ and
$H^2=\frac{1}{\kappa^2z}$. Moreover we can see from the system
(\ref{3.4.1})-(\ref{3.4.2}) that the fixed point $x_0=1,\,\,z_0=0$
always exists, except for some very special choices of functions $f$
and $G$.

It is interesting to note that actually due to the constraint
equation (\ref{3.1}), in the case at hand, there is only one
independent variable, but by introducing the variable $z$ and by
taking into account the equation (\ref{3.1.1}), allows us to obtain
more information about the evolution of the dynamical system. For
instance, one of the most hard tasks in such kind of dynamical
systems analysis, is to interpret correctly the physical meaning of
the stationary points. This approach was firstly proposed in Ref.
\cite{BH}. Let us introduce an additional parameter, which will be
very helpful for this interpretation, which is the effective
equation of state, which we denote as $w_{eff}$, and it is defined
as follows:
\begin{equation}
w_{eff}\equiv -1-\frac{2\dot H}{3 H^2}. \label{3.4.3}
\end{equation}
The parameter $w_{eff}$ can be expressed in terms of the
dimensionless variables (\ref{3.1})-(\ref{3.1.1}) as follows,
 \begin{equation}
w_{eff}=\frac{1}{3}-\frac{4}{3}x+\frac{1}{\rho_{tot}}(f+G).
\label{3.4.4}
\end{equation}
By specifying the EoS, it is possible to obtain various physically
interesting evolution scenarios, so in the rest of this section we
shall specify the EoS and we study in detail the dynamical evolution
stemming from the choice of EoS.


\subsection{A Simplified Form of EoS}

Let us discuss the simplest case of EoS, which is $p=w_0\rho
+w_1H^2$, which in turn implies $f(\rho)=\rho(1+w_0)$,
$G(H)=w_1H^2$. It is easy to verify that in this case, the equations
(\ref{3.4.1})-(\ref{3.4.2}) take the following form,
\begin{eqnarray}
x'&&= x^2(3w_0-1) +x(1-3w_0+w_1\kappa^2)-w_1\kappa^2,\label{3.5}\\
z'&&= z\left[ x(3w_0-1) +4+w_1\kappa^2\right].\label{3.6}
\end{eqnarray}
Thus we have the following stationary points for the dynamical
system above,
\begin{itemize}
\item First fixed point  $x_0=\frac{w_1\kappa^2}{1-3w_0},\,\,\,z_0=0$.
\item Second fixed point $x_0=1,\,\,\,z_0=0$.
\end{itemize}

\begin{figure}[h!]
  \includegraphics[width=10cm]{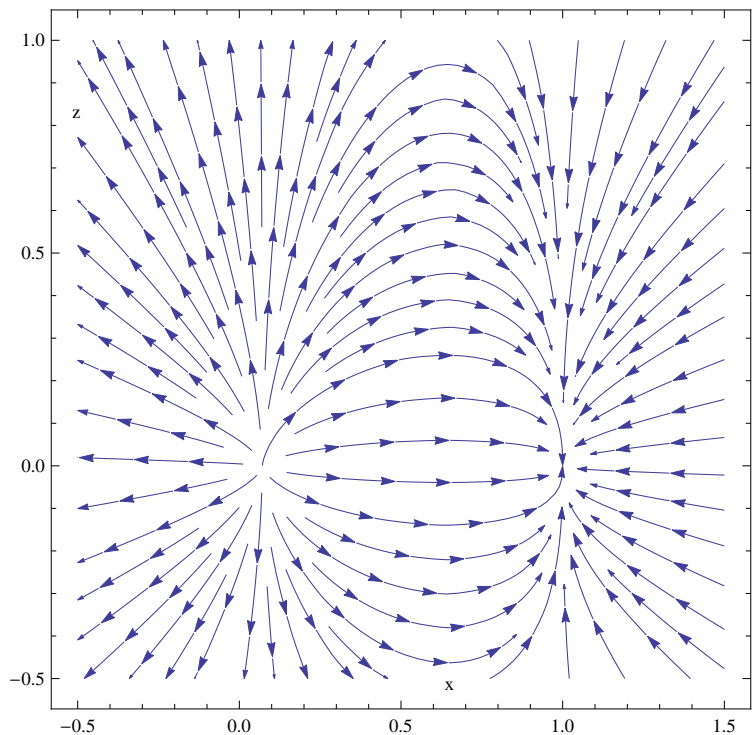}
  \caption{Phase portrait for the case $w_0=-2$, $w_1\kappa^2=0.5$.
   $w_{eff}=\frac{1}{3}$ for the left point and $w_{eff}=-\frac{11}{6}$ for the right point.}
  \label{fig:fig4}
\end{figure}

\begin{figure}[h!]
  \includegraphics[width=10cm]{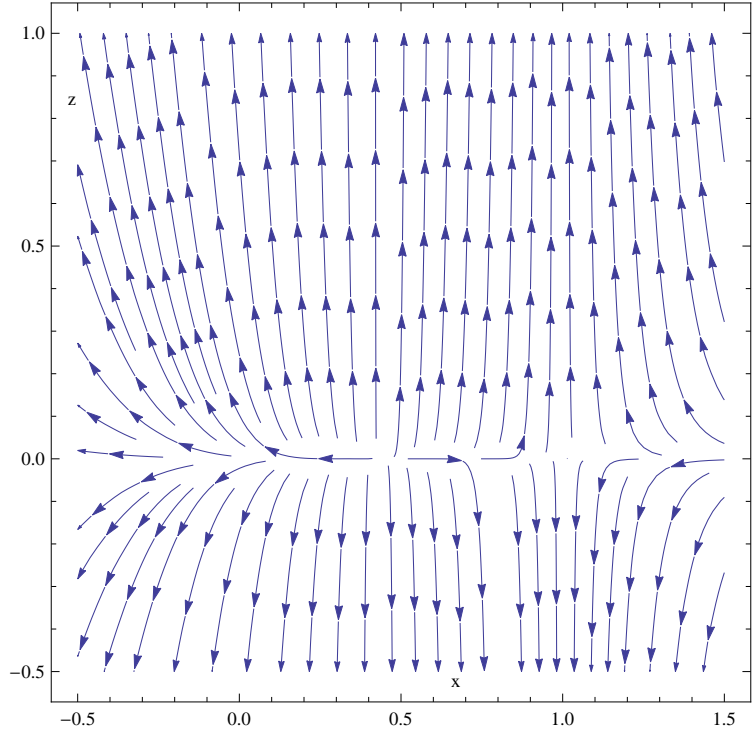}
  \caption{Phase portrait for the case $w_0=0.1$, $w_1\kappa^2=0.3$.  $w_{eff}=\frac{1}{3}$ for the left point and $w_{eff}=0.2$ for the right point.}
  \label{fig:fig5}
\end{figure}

For the first fixed point we need to note that, this point lies in
the physical region only if $0<\frac{w_1\kappa^2}{1-3w_0}<1$.
The corresponding eigenvalues are $\mu_1=1-3w_0-w_1\kappa^2$, and
$\mu_2=4$. Using Eq. (\ref{3.4.4}) we find that at this point we
have $w_{eff}=\frac{1}{3}$.
With regard to the second fixed point,
the eigenvalues are $\mu_1=3w_0+w_1\kappa^2-1$, and $\mu_2=\mu_1+4$.
Correspondingly we find that at this point
$w_{eff}=w_0+\frac{1}{3}w_1\kappa^2$. Thus we can see that the first
fixed point always unstable and also that the second fixed point may
be stable only if one (or both) of parameters $w_0$, $w_1$ are
strictly negative. For instance if we require that the first point
lies in the physical region, we find that the second point is stable
for $w_0<-1$. In Figs.  (\ref{fig:fig4}-\ref{fig:fig5}) we plot the
typical behavior of the phase trajectories.

Particularly, Fig. \ref{fig:fig4} corresponds to $w_0=-2$, and
$w_1\kappa^2=0.5$. The left fixed point corresponds to the effective
EoS parameter $w_{eff}=\frac{1}{3}$ and for the right fixed point,
we have $w_{eff}=-\frac{11}{6}$. Clearly the left fixed point
represents radiation, while the right one corresponds to some
phantom evolution. In Fig. \ref{fig:fig5}, the phase portrait
corresponds to the following choices for the parameters, $w_0=0.1$,
$w_1\kappa^2=0.3$. The left point corresponds to an effective EoS
parameter $w_{eff}=\frac{1}{3}$, while the right point corresponds
to $w_{eff}=0.2$, which describes a form of collisional matter
\cite{Oikonomou:2014lua}.

\subsection{More Complicated Forms of the EoS}

Now let us study more complicated forms of the EoS, and we choose it
to be of the form $f(\rho)+G(H)=A\rho^{\alpha}+BH^{2\beta}$. This
EoS is known to lead the cosmological system to finite-time
singularities, as this was demonstrated in Ref. \cite{NO1}. For this
EoS, the equations (\ref{3.4.1})-(\ref{3.4.2}) take the form,
\begin{eqnarray}
x'&&= -4x^2+4x+\kappa^4z\left[ A\left(\frac{3x}{\kappa^4z}\right)^{\alpha}+B\left(\frac{1}{\kappa^2z}\right)^{\beta} \right](x-1),\label{3.7}\\
z'&&=4z(1-x)+\kappa^4z^2\left[ A\left(\frac{3x}{\kappa^4z}\right)^{\alpha}+B\left(\frac{1}{\kappa^2z}\right)^{\beta} \right].\label{3.8}
\end{eqnarray}
Correspondingly, the effective EoS in this case reads,
 \begin{equation}
w_{eff}=\frac{1}{3}\left [1-4x + A\kappa^4z\left(\frac{3x}{\kappa^4 z}\right)^{\alpha}+B\kappa^4 z\frac{1}{(\kappa^2 z)^{\beta}}\right ].
\label{3.8.1}
\end{equation}

It is clear that in the most general case, this system may be solved
only numerically, so let study some appropriately chosen cases which
admit analytical solutions. Consider first the case for which
$\alpha=1$, $\beta=2$, in which case, the dynamical system
(\ref{3.7})-(\ref{3.8}) takes the following form,
\begin{eqnarray}
x'&&= \left( 3Ax-4x+\frac{B}{z}\right)(x-1),\label{3.9}\\
z'&&=4z(1-x)+3Axz+B.\label{3.10}
\end{eqnarray}
In this case there are two stationary points: the first is $x_0=1$,
$z_0=0$\footnote{Note here that in all these three cases existence
of the point $x_0=1$, $z_0=0$ is not an obvious solution, but it can
be confirmed by numerical investigations.} and the second is
$x_0=1$, $z_0=\frac{-B}{3A}$. For the first point, we have
$w_{eff}=sign(B)\infty$\footnote{Infinite values of $w_{eff}$ looks
like the Ruzmaikin solution at $t\rightarrow 0$.} and for the second
one we have $w_{eff}=-1$.

\begin{figure}[h!]
  \includegraphics[width=10cm]{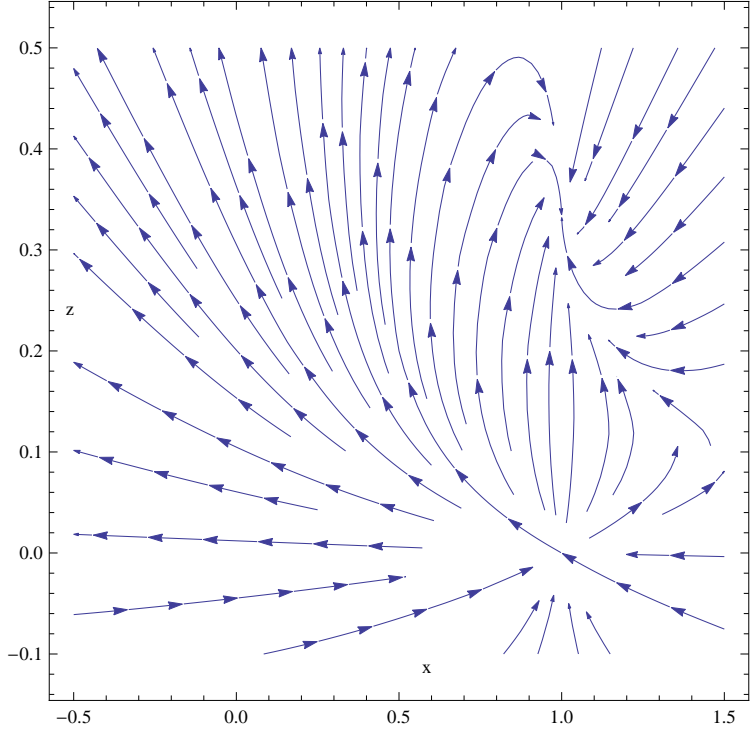}
  \caption{Phase portrait for the case $A=-1$, $B=1$, $\alpha=1$, and $\beta=2$.}
  \label{fig:fig6}
\end{figure}

We can see that the second point corresponds to some new non-trivial
de Sitter state with $\rho_{tot}=\rho$ and $H^2=H_0^2\neq 0$. The
eigenvalues of the second fixed point are $\mu_1=-4$, and
$\mu_2=3A$. The typical behavior of the phase trajectories
corresponding to this case, can be found in Fig.\ref{fig:fig6}, for
$A=-1$, and $B=1$. As it can be seen, there exist trajectories which
start from the first fixed point  $x_0=1$, $z_0=0$ (note that zero
values of $z$ correspond to infinite values of $H$, which indicates
a singularity) and end up to the second fixed point, with
non-singular and non-zero values of $H$. In effect, the second fixed
point may be interpreted as a late-time acceleration de Sitter point
of the cosmological dynamical system.

\begin{figure}[h!]
  \includegraphics[width=10cm]{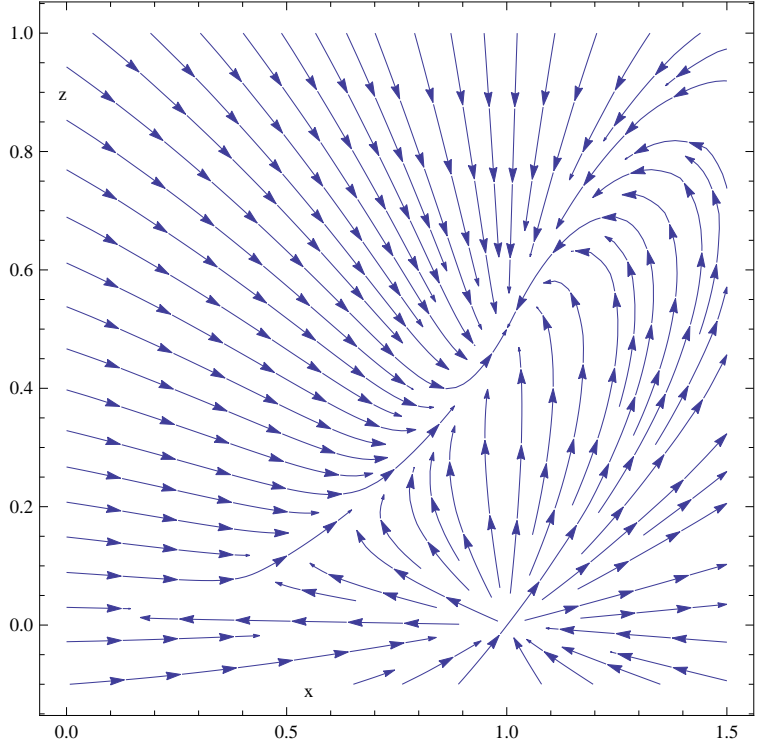}
  \caption{Phase portrait for the case $A=\frac{1}{2}$, $B=-9$, $\kappa^2=1$, $\alpha=2$, and $\beta=1$.}
  \label{fig:fig7}
\end{figure}

Now let us consider the case for which $\alpha=2$, and $\beta=1$, in
which case, the dynamical system (\ref{3.7})-(\ref{3.8}) takes the
form,
\begin{eqnarray}
x'&&= \left( \frac{9A}{\kappa^4}\frac{x^2}{z}-4x+\kappa^2 B\right)(x-1),\label{3.11}\\
z'&&=4z(1-x)+\frac{9A}{\kappa^4}x^2+\kappa^2 B z.\label{3.12}
\end{eqnarray}
In this case there are two stationary points, namely, $x_0=1$,
$z_0=0$ and also $x_0=1$, $z_0=\frac{-9A}{\kappa^6B}$ and the
situation is very similar to the previous case. For the first fixed
point we have $w_{eff}=sign(A)\infty$ and for the second one
$w_{eff}=-1$. The eigenvalues of the second point are $\mu_1=-4$,
and $\mu_2=\kappa^2B$. Note also that for both these cases, $A$ and
$B$ must have opposite sign, in order for the second fixed point to
lie in the physical region. We also need to note that for this case
there are additional fixed points, which are difficult to find
analytically, but the most physically interesting cases of fixed
points are the ones we just presented. The phase space behavior
corresponding to this case can be found in Fig. \ref{fig:fig7}, for
$A=\frac{1}{2}$, $B=-9$, $\kappa^2=1$, and as it can be seen, the
behavior of the trajectories is similar to the previous case.

\begin{figure}[h!]
  \includegraphics[width=10cm]{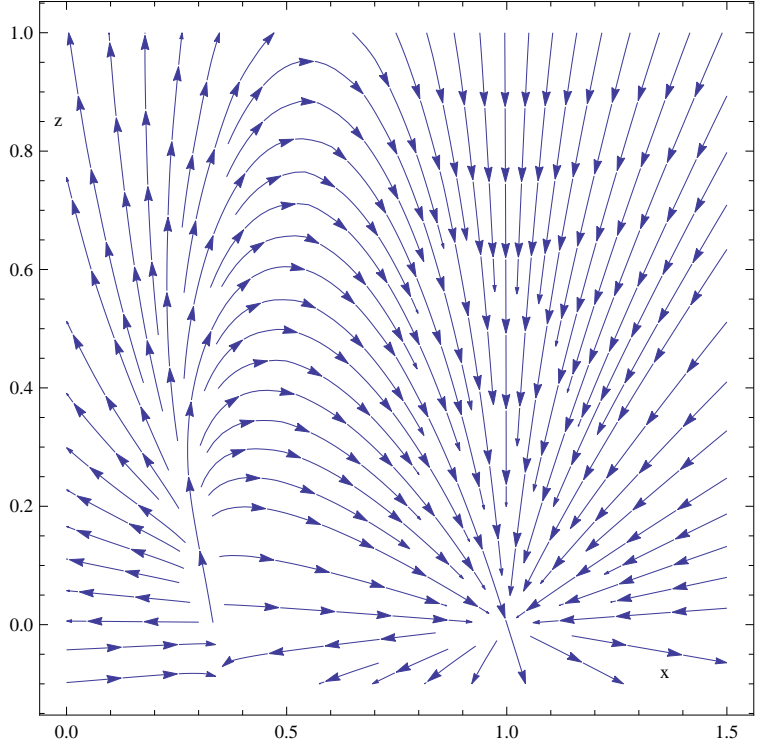}
  \caption{Phase portrait for the case $A=-1$, $B=1$, $\kappa^2=1$}
  \label{fig:fig8}
\end{figure}

\begin{figure}[h!]
  \includegraphics[width=10cm]{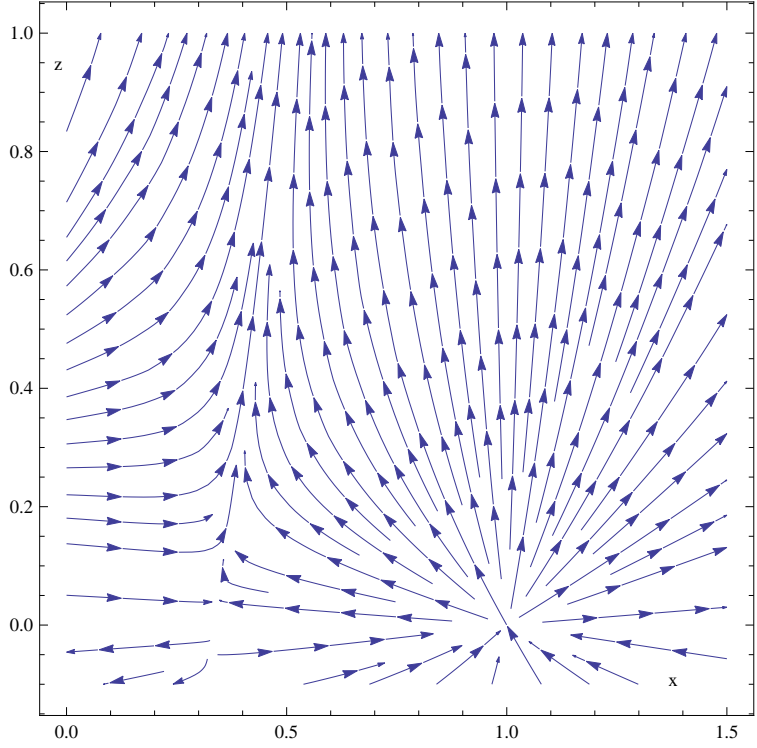}
  \caption{Phase portrait for the case $A=1$, $B=-1$, $\kappa^2=1$}
  \label{fig:fig9}
\end{figure}

Concluding this section, let us briefly discuss another choice of
parameters for which it is possible to find analytically the fixed
points, and this occurs for the choice $\alpha=2$, $\beta=2$, in
which case, the dynamical system (\ref{3.7})-(\ref{3.8}) takes the
form,
\begin{eqnarray}
x'&&= \left( \frac{9A}{\kappa^4}\frac{x^2}{z}-4x+\frac{B}{z}\right)(x-1),\label{3.13}\\
z'&&=4z(1-x)+\frac{9A}{\kappa^4}x^2+B.\label{3.14}
\end{eqnarray}
In this case, the stationary points are the following, $x_0=1$,
$z=0$ and $x_0^2=\frac{-B\kappa^4}{9A}$, $z=0$. For the first fixed
point we have $w_{eff}=sign(B\kappa^4+9A)\infty$ and for the second
one $w_{eff}=\frac{1}{3}-\frac{4}{9}\kappa^2\sqrt{\frac{-B}{A}}$.
The behavior of the trajectories corresponding to this case can be
found in Fig. \ref{fig:fig8}, for the choice $A=-1$, $B=1$, and in
Fig. \ref{fig:fig9}, for the choice $A=1$, $B=-1$.

\section{Dark Matter Interacting with Dark Energy: Some non-trivial Models}\label{sec:4}

Let us now discuss some non-trivial models which describe
interactions between the dark matter and dark energy fluids. The
dynamical system in this case may be written in the following form,
\begin{eqnarray}
\frac{3}{\kappa^2}H^2&&=\rho_r+\rho+\rho_{dm}=\rho_{tot},\label{4.1}\\
\frac{-2}{\kappa^2}\dot H &&= \frac{4}{3}\rho_r + [f(\rho)+G(H)]+\rho_{dm}+(-3H\zeta),\label{4.1.1}\\
\dot\rho_r&&=-4H\rho_r,\label{4.2}\\
\dot\rho &&=-3H\left[f(\rho)+G(H)+\frac{Q}{3H}H^{2k}\right],\label{4.3}\\
\dot\rho_{dm}&&=-3H\left[\rho_{dm}-\frac{Q}{3H}H^{2k}+(-3H\zeta)\right],\label{4.4}
\end{eqnarray}
where we modified the interaction between dark energy and dark
matter by using the multiplier $H^{2k}$\footnote{Note that case
$k=0$ corresponds to the standard interaction we presented in a
previous section.}, and also the term $-3H\zeta$, which corresponds
to the bulk viscous pressure of the dark matter fluid. We will
assume that the bulk viscous coefficient $\zeta$ has the following
form,
\begin{equation}
\zeta=\frac{\zeta_0}{\sqrt{3\kappa^2}}\rho_{tot}^{\frac{1}{2}}=\frac{1}{\kappa^2}H\zeta_0.
\label{4.5}
\end{equation}
We define the set of dimensionless variables as follows,
\begin{eqnarray}
x&&=\frac{\kappa^2}{3H^2}\rho=\frac{\rho}{\rho_{tot}},\label{4.6}\\
y&&=\frac{\kappa^2}{3H^2}\rho_{dm}=\frac{\rho_{dm}}{\rho_{tot}},\label{4.7}\\
z&&=\frac{1}{\kappa^2H^2}=\frac{3}{\rho_{tot}\kappa^4},\label{4.7.1}\\
q&&=\frac{\kappa^2}{3H^3}Q=\frac{Q}{H\rho_{tot}},\label{4.8}
\end{eqnarray}
where only $x$ and $y$ are dynamical independent variables. In the
new variables system, Eqs. (\ref{4.1})-(\ref{4.4}) take the
form\footnote{Note that the constraint (\ref{4.1}) and the equation
for $\rho_r$ have already been taken into account for this system.}:
\begin{eqnarray}
x'&&=-4x^2+4x-yx-3x\zeta_0-\frac{q}{(\kappa^2z)^k}+\kappa^4z(f+G)(x-1),\label{4.9}\\
y'&&=-y^2-4xy+y(1-3\zeta_0)+3\zeta_0+\frac{q}{(\kappa^2z)^k}+\kappa^4zy(f+G),\label{4.10}\\
z'&&=4z(1-x-y)+z^2\kappa^4(f+G)+3yz-3z\zeta_0,\label{4.10.1}
\end{eqnarray}
and using that the EoS has the form we used in the previous section,
namely, $f(\rho)=\rho(1+w_0)$, $G(H)=w_1H^2$, we obtain,
\begin{eqnarray}
x'&&=(x-1)(w_1\kappa^2+3xw_0-x)-x(y+3\zeta_0)-\frac{q}{(\kappa^2z)^k},\label{4.11}\\
y'&&=-y^2+y(1-3\zeta_0+3xw_0-x+w_1\kappa^2)+3\zeta_0+\frac{q}{(\kappa^2z)^k},\label{4.12}\\
z'&&=z\left(4-x-y+3xw_0-3\zeta_0+w_1\kappa^2\right).\label{4.13}
\end{eqnarray}
The effective EoS in this case (for arbitrary functions $f$ and $G$)
reads,
 \begin{equation}
w_{eff}=\frac{1}{3}\left [1-4x -y + \kappa^4z\left(f+G\right) -3\zeta_0 \right ],
\label{4.13.1}
\end{equation}
which in the case that $f(\rho)=\rho(1+w_0)$, $G(H)=w_1H^2$,
becomes,
 \begin{equation}
w_{eff}=\frac{1}{3}\left [1-x -y + 3xw_0 +w_1\kappa^2 -3\zeta_0 \right ].
\label{4.13.2}
\end{equation}
Now we consider some special cases, and we start with the case
$k=0$. This case corresponds to the usual interaction term. Since
the first two  equations are identical to (\ref{2.11})-(\ref{2.12}),
we have the solutions (\ref{2.11.0.0}), (\ref{2.11.0.1}), where we
need to add $z_0=0$.
Also, since the first two equations do not depend on $z$, the first
and second eigenvalues will be totally identical to the ones we
obtained in section \ref{sec:2}, and one additional eigenvalue
appears for every stationary point, which is,
\begin{equation}
\mu_3=4-x_0-y_0+3x_0w_0-3\zeta_0+w_1\kappa^2.
\label{4.14}
\end{equation}
Also note that the stability or instability of the point $z_0=0$
with respect to this additional coordinate $z$, implies stability in
the past or in the future correspondingly. It mean that a stable
with respect to coordinates $x$, $y$ point which have $z_0=0$ will
be stable in the past if it is stable with respect to coordinate
$z$, or equivalently will be stable for infinite values of $H$. In
addition, it will stable in the future, if it is unstable with
respect to coordinate $z$, or equivalently it will be stable for
small (zero) values of $H$.

The effective EoS (\ref{4.13.2}) for the fixed point
(\ref{2.11.0.0}) is equal to $w_{eff}=\frac{1}{3}$ for any values of
parameters, which clearly describes a radiation dominated era.
However, the effective EoS for the fixed points (\ref{2.11.0.1}),
has a more complicate structure, which is given below,
\begin{equation}
w_{eff}=\frac{1}{2}w_0+\frac{1}{6}w_1\kappa^2-\frac{1}{2}\zeta_0 \pm \frac{1}{6}\sqrt{(3w_0+w_1\kappa^2-3\zeta_0)^2+12w_0q+36w_0\zeta_0}.
\label{4.14.1}
\end{equation}
Let us now further analyze the case at hand, by specifying the
values of the free parameters, so we start with the parameter $w_0$,
and assume for the moment that $w_0=0$. In this case the fixed point
(\ref{2.12.0}) has an additional eigenvalue $\mu_3=4$ and the
corresponding effective EoS becomes $w_{eff}=\frac{1}{3}$, which
describes radiation. Moreover, the fixed point (\ref{2.15}) has the
additional eigenvalue $\mu_3=3-3\zeta_0+w_1\kappa^2$ and the
corresponding effective EoS for this point is
$w_{eff}=\frac{1}{3}w_1\kappa^2-\zeta_0$. Thus in Fig.
\ref{fig:fig0} the left fixed point corresponds to radiation with
$w_{eff}=\frac{1}{3}$, and the right point has
$w_{eff}=\frac{11}{3}$, and for both the fixed points, $H$ has
infinite values. In Table \ref{tab1} we have gathered all the fixed
points which correspond to the case $k=0$, $w_0=0$.
\begin{table}[H]
\caption{\label{tab1}Fixed Points for the Case $k=0$, $w_0=0$.}
\begin{center}
\begin{tabular}{|c|c|c|c|c|c|c|c|}
\hline
$P_i$ & $x_0$           & $y_0$                                     & $z_0$ & $\mu_1$                                           & $\mu_2$                  & $\mu_3$ & $w_{eff}$ \\
\hline
  1a  & $q+w_1\kappa^2$ & $-3\zeta_0-q$                             &   $0$ &   $1$                                             & $1-w_1\kappa^2+3\zeta_0$ & $4$     &   $\frac{1}{3}$       \\
\hline
  1b  & $1-y_0$         & $\frac{3\zeta_0+q}{3\zeta_0-w_1\kappa^2}$ &   $0$ & $w_1\kappa^2-3\zeta_0$ & $w_1\kappa^2-3\zeta_0-1$ & $3-3\zeta_0+w_1\kappa^2$ & $\frac{1}{3}w_1\kappa^2-\zeta_0$  \\
\hline

\end{tabular}
\end{center}
\end{table}
As it can be seen in Table \ref{tab1}, the fixed point
$P_{1\mathrm{b}}$ may describe late-time acceleration. Indeed, if we
set $w_{eff}=-1+\alpha$, with $0<\alpha\ll 1$, we obtain
$\mu_1=-3+3\alpha$, $\mu_2=-4+3\alpha$ and $\mu_3=\alpha$ and as we
already noted, this means that this point is stable in the future
(for small values of $H$). Moreover, by changing the parameter $q$,
we can provide any interesting relation between $\rho_{dm}$ and
$\rho$. For instance if put $q=\frac{3}{4}-3\zeta_0$ we obtain for
this point $y_0\equiv\frac{\rho_{dm}}{\rho_{tot}}=\frac{1}{4}$.

Let us discuss some alternative choices for the parameters, so
consider the case $q=-3\zeta_0$, in which case the fixed point
(\ref{2.15.0}) has an additional eigenvalue, which we denote
$\mu_3$, and it is equal to,
$\mu_3=4-\frac{3\zeta_0-w_1\kappa^2}{3w_0}$ and the corresponding
effective EoS is $w_{eff}=\frac{1}{3}$. In addition, the fixed point
(\ref{2.15.1}) has the additional eigenvalue
$\mu_3=3+3w_0-3\zeta_0+w_1\kappa^2$ with $w_{eff}=0$, which
describes a matter dominated state. Finally, the fixed point
(\ref{2.16}) has $\mu_3=3$ with
$w_{eff}=w_0+\frac{1}{3}w_1\kappa^2-\zeta_0$. In Table \ref{tab2} we
have gathered all the fixed points for the case $q=-3\zeta_0$.
\begin{table}[H]
\caption{\label{tab2}Fixed Points for the Case $q=-3\zeta_0$.}
\begin{center}
\begin{tabular}{|c|c|c|c|c|c|c|c|}
\hline
$P_i$ & $x_0$           & $y_0$                                     & $z_0$ & $\mu_1$                                           & $\mu_2$                  & $\mu_3$ & $w_{eff}$ \\
\hline
  2a  & $\frac{3\zeta_0-w_1\kappa^2}{3w_0} $ & $0$                             & $0$ &   $1-x_0$                                             & $2\mu_1-w_1\kappa^2+3\zeta_0-3w_0-1$ & $4-\frac{3\zeta_0-w_1\kappa^2}{3w_0}$     &   $\frac{1}{3}$       \\
\hline
  2b  & $1$         & $0$ &  $0$ & $w_1\kappa^2-3\zeta_0+3w_0$ &  $w_1\kappa^2-3\zeta_0+3w_0-1$ & $3+3w_0-3\zeta_0+w_1\kappa^2$ & $0$  \\
\hline
  2c  & $1-y_0$         & $\frac{3w_0-3\zeta_0+w_1\kappa^2}{3w_0}$ & $0$ & $-1$ & $-3w_0-w_1\kappa^2+3\zeta_0$ & $3$ & $w_0+\frac{1}{3}w_1\kappa^2-\zeta_0$  \\
\hline

\end{tabular}
\end{center}
\end{table}
As it can be seen in Table \ref{tab2}, the fixed point $P_{2a}$ is
always unstable, if it lies in the physical region, that is, when
$x_0\leqslant 1$. The fixed point $P_{2b}$ may be stable in the past
and in the future, depending on the values of the parameters.
Finally, the fixed point $P_{2c}$ may be stable only if $w_{eff}>0$.
So in this case, no fixed point describes late-time acceleration,
however, the fixed points $P_{2a}$ and $P_{2b}$ may describe a
radiation dominated era and matter dominated era respectively.

Consider now the case $q=-3\zeta_0$, $q=-w_1\kappa^2$, in which
case, the fixed point (\ref{2.17}) has the additional eigenvalue
$\mu_3=3$ and the corresponding effective EoS is $w_{eff}=w_0$.
Accordingly, the fixed point (\ref{2.18}) has $\mu_3=3+3w_0$ with
$w_{eff}=0$ and finally the fixed point (\ref{2.19}) has $\mu_3=4$
with $w_{eff}=\frac{1}{3}$. Thus in Fig.\ref{fig:fig1} we have
$w_{eff}=\frac{1}{3}$ for the fixed point $(x_0=0,y_0=0)$,
$w_{eff}=0$ for the fixed point $(x_0=1,y_0=0)$ and
$w_{eff}=\frac{1}{2}$ for the fixed point $(x_0=0,y_0=1)$.
Correspondingly, in Fig. \ref{fig:fig2} we have
$w_{eff}=\frac{1}{3}$ for the fixed point $(x_0=0,y_0=0)$,
$w_{eff}=0$ for the fixed point $(x_0=1,y_0=0)$ and
$w_{eff}=-\frac{1}{2}$ for the fixed point $(x_0=0,y_0=1)$. Note
that all the aforementioned fixed points correspond to states which
have infinite values of $H$.
\begin{table}[H]
\caption{\label{tab3}Fixed Points for  the Case $q=-3\zeta_0$,
$q=-w_1\kappa^2$.}
\begin{center}
\begin{tabular}{|c|c|c|c|c|c|c|c|}
\hline
$P_i$ & $x_0$ & $y_0$ & $z_0$ & $\mu_1$ &   $\mu_2$ & $\mu_3$ & $w_{eff}$ \\
\hline
  3a  & $0$   & $0$   & $0$   & $1$     & $1-3w_0$  & $4$     & $\frac{1}{3}$  \\
\hline
  3b  & $1$   & $0$   & $0$   & $3w_0$  & $3w_0-1$ & $3+3w_0$ & $0$  \\
\hline
  3c  & $0$   & $1$   & $0$   &   $-1$  &  $-3w_0$ & $3$      &   $w_0$       \\
\hline

\end{tabular}
\end{center}
\end{table}
In Table \ref{tab3} we have gathered all the fixed points for the
case $q=-3\zeta_0$, $q=-w_1\kappa^2$.

Now let us consider some alternative choices for the parameter $k$,
and we start with the case $k=1$, in which case the only stationary
point is the following,
\begin{equation}
x_0=\frac{3\zeta_0-3-w_1\kappa^2}{3w_0},\,\,\,y_0=1-x_0,\,\,\,z_0=\frac{qw_0}{\kappa^2(3+3w_0+w_1\kappa^2-3\zeta_0-3w_0\zeta_0)},
\label{4.15}
\end{equation}
and the expression (\ref{4.13.2}) yields for this fixed point,
\begin{equation}
w_{eff}=-1. \label{4.15.1}
\end{equation}
The effective EoS parameter above describes a de Sitter evolution,
with a finite value for the Hubble parameter, namely $H=H_0$, which
may be arbitrarily small, depending on the values of the other
parameters. The eigenvalues for this fixed point in the most general
case, have a quite complicate form, but there is one special case
for which we can compute these analytically, and this occurs when
$w_0=\frac{1}{3}$. In this case, the eigenvalues are,
\begin{eqnarray}
\mu_1&&=-4,\label{4.15.2}\\
\mu_{2,3}&&=\frac{1}{2}\left[ 3\zeta_0-7-w_1\kappa^2\pm \sqrt{(3\zeta_0-7-w_1\kappa^2)^2+12(4-4\zeta_0+w_1\kappa^2)} \right ],\label{4.15.3}\\
\end{eqnarray}
where the first eigenvalue corresponds to variable $x$. We can see
that the stability with respect to all dimensions, requires the
following conditions to hold true,
\begin{eqnarray}
\zeta_0&&<\frac{7}{3}+\frac{1}{3}w_1\kappa^2,\label{4.15.4}\\
\zeta_0&&>1+\frac{1}{4}w_1\kappa^2,\label{4.15.5}
\end{eqnarray}
which is quite compatible with $0\leqslant x_0\leqslant 1$, but
incompatible with $z_0>0$, which reads
$\zeta_0<1+\frac{1}{4}w_1\kappa^2$. This means that by varying the
parameters, we can make this point stable with respect to the
coordinates $x$ and $y$, and unstable with respect $z$, so this
point may be used for the construction of the late-time acceleration
phase. Note also that by changing value of $w_0$, it is quite
possible to make this point stable with respect to all the
coordinates.

Consider now the case $k=-1$, in which case there are four
stationary points, which can be found in Table \ref{tab4}. Note that
the parameter $a$ appearing in Table \ref{tab4} is equal to
$a=(3w_0-w_1\kappa^2+3\zeta_0)^2+12\kappa^2w_0w_1$. The
corresponding eigenvalues can be found in Table \ref{tab4a}.
\begin{table}[H]
\caption{\label{tab4} Fixed points for the Case $k=-1$.}
\begin{center}
\begin{tabular}{|c|c|c|c|c|}
\hline
$P_i$ & $x_0$ & $y_0$ & $z_0$ & $w_{eff}$ \\
\hline
  4a  & $\frac{w_1\kappa^2}{1-3w_0}$   & $-3\zeta_0$   & $0$ & $\frac{1}{3}$  \\
\hline
  4b  & $\frac{3\zeta_0-3-w_1\kappa^2}{3w_0}$   & $1-x_0$ & $\frac{3+3w_0+w_1\kappa^2-3\zeta_0-3w_0\zeta_0}{\kappa^2qw_0}$  & $-1$  \\
\hline
  4c  & $\frac{1}{6w_0}\left[3w_0-w_1\kappa^2+3\zeta_0+\sqrt{a} \right]$   & $1-x_0$   &  $0$  &    $\frac{1}{2}w_0+\frac{1}{6}w_1\kappa^2-\frac{1}{2}\zeta_0+\sqrt{a}$       \\
\hline
  4d  & $\frac{1}{6w_0}\left[3w_0-w_1\kappa^2+3\zeta_0-\sqrt{a} \right]$   & $1-x_0$   &  $0$  &   $\frac{1}{2}w_0+\frac{1}{6}w_1\kappa^2-\frac{1}{2}\zeta_0-\sqrt{a}$       \\
\hline

\end{tabular}
\end{center}
\end{table}
By looking the Table \ref{tab4} it can be seen that the most
phenomenologically interesting fixed point is $P_{4b}$. First of all
it is easy to make this fixed point a stable attractor. Moreover
$w_{eff}$ is equal to $-1$ exactly, so this point corresponds to
some de Sitter solution. Finally, for this point we have $z_0\neq 0$
which means that this point corresponds to some state with non-zero
(and non-infinite) value of the Hubble parameter $H=H_0$. So by
changing the values of the parameters it is easy to make $z_0$
sufficiently large, which implies sufficiently small values of
$H_0$. For instance, by choosing $w_0>0$, $w_1>0$ and $\zeta_0<0$ we
can make this fixed point stable. By assigning sufficiently large
values of $w_1$ for sufficiently small (positive) values of $q$, we
get large values for $z_0$. And finally we can see that by changing
the values of the parameters, it is easy to construct some fixed
relation $\frac{y_0}{x_0}\equiv \frac{\rho_{dm}}{\rho}$. In
conclusion, this de Sitter attractor may be viewed as a late-time
attractor.
\begin{table}[H]
\caption{\label{tab4a}Eigenvalues for the case $k=-1$.}
\begin{center}
\begin{tabular}{|c|c|c|c|}
\hline
$P_i$ & $\mu_1$ & $\mu_2$ & $\mu_3$ \\
\hline
  4a  & $1$   & $1-3w_0-w_1\kappa^2+3\zeta_0$   & $4$   \\
\hline
  4b  & $-3$    & $-(3+3w_0+w_1\kappa^2-3\zeta_0)$ & $-4$  \\
\hline
  4c  & $-1+w_1\kappa^2+3w_0x_0-3\zeta_0$   & $w_1\kappa^2+6w_0x_0-3\zeta_0-3w_0$   &  $\mu_1+4$ \\
\hline
  4d  & $-1+w_1\kappa^2+3w_0x_0-3\zeta_0$   & $w_1\kappa^2+6w_0x_0-3\zeta_0-3w_0$   &  $\mu_1+4$   \\
\hline

\end{tabular}
\end{center}
\end{table}
It is interesting to note that there are five different regions in
the parameter space, for which all four stationary points lie in the
physical region, and these regions are the following,\footnote{i.e.
$0\leqslant x_0\leqslant 1$, $0\leqslant y_0\leqslant 1$,
$z\geqslant0$ for all points simultaneously.}
\begin{eqnarray}
w_1\kappa^2\leqslant-4,\,\,q>0,\,\, -\frac{1}{3}\leqslant\zeta_0<a,\,\,w_0\geqslant b,\label{4.19}\\
w_1\kappa^2\leqslant-4,\,\,q>0,\,\, a\leqslant\zeta_0\leqslant0,\,\,w_0\geqslant c,\label{4.20}\\
-4<w_1\kappa^2<-\frac{2}{3}\sqrt{13}-\frac{2}{3},\,\,q>0,\,\, -c+\frac{4}{3}\leqslant\zeta_0<a,\,\,w_0\geqslant b,\label{4.21}\\
-4<w_1\kappa^2<-\frac{2}{3}\sqrt{13}-\frac{2}{3},\,\,q>0,\,\, a\leqslant\zeta_0<0,\,\,w_0\geqslant c,\label{4.22}\\
-\frac{2}{3}\sqrt{13}-\frac{2}{3}\leqslant w_1\kappa^2\leqslant-3,\,\,q>0,\,\, -c+\frac{4}{3}\leqslant\zeta_0\leqslant 0,\,\,w_0\geqslant c,\label{4.23}
\end{eqnarray}
where
$$
a=\frac{1}{3}(2w_1\kappa^2-1)+\frac{2}{3}\sqrt{w_1^2\kappa^4-w_1\kappa^2},
$$

$$
b=-\frac{1}{3}w_1\kappa^2-\zeta_0+\frac{2}{3}\sqrt{3w_1\kappa^2\zeta_0},
$$

$$
c=\frac{1}{3}(1-w_1\kappa^2).
$$
However, the above cases are quite hard to be tackled analytically,
so a numerical study is needed, which exceeds the purposes of this
article.

\subsection{Oscillating Dark Energy-Dark Matter Interaction}

In the previous sections we considered cases of dark energy and dark
matter interactions, by specifying the free parameters, and in this
section we follow a different approach: we will directly modify the
functional form of the dark energy-dark matter interaction by making
it oscillating. So assume that in Eq. (\ref{4.3}) we make the
following replacement, $-QH^{2k} \rightarrow -Q \cos(h_0H^2)$. In
effect, instead of Eqs. (\ref{4.11})-(\ref{4.13}) we have
\begin{eqnarray}
x'&&=(x-1)(w_1\kappa^2+3xw_0-x)-x(y+3\zeta_0)-q\cos(h_0H^2),\label{4.24}\\
y'&&=-y^2+y(1-3\zeta_0+3xw_0-x+w_1\kappa^2)+3\zeta_0+q\cos(h_0H^2),\label{4.25}\\
z'&&=z\left(4-x-y+3xw_0-3\zeta_0+w_1\kappa^2\right),\label{4.26}
\end{eqnarray}
where $\cos(h_0H^2)=\cos\left (\frac{h_0}{\kappa^2 z} \right )$. Let
us suppose that the argument of the cosine function changes
monotonically, from $\frac{\pi}{2}$ during the inflationary era,
when $H^2\thicksim M_{Pl}$, to $0$ which corresponds at the present
time. In practice we may realize this kind of behavior by assuming
sufficiently small values of the parameter $h_0$. Then, the physical
interpretation of the situation at hand is the following, at early
times (inflationary era), there is no interaction between the
various matter fluid components, and as the time grows, this
interaction starts to develop, and it grows until it reaches its
maximum  value at $H=0$. From a mathematical point of view, this
would mean that there are two distinct asymptotic states at
$t\rightarrow 0$ and $t\rightarrow\infty$, which are described by
two distinct dynamical systems. In the following we discuss these
two asymptotic dynamical systems separately.


\subsubsection{Asymptotic State I: The Inflationary Epoch}\label{infl}

Let us now focus on the inflationary asymptotic state, in which case
instead of (\ref{4.24})-(\ref{4.26}), we have the following
dynamical system,
\begin{eqnarray}
x'&&=(x-1)(w_1\kappa^2+3xw_0-x)-x(y+3\zeta_0),\label{4.27}\\
y'&&=-y^2+y(1-3\zeta_0+3xw_0-x+w_1\kappa^2)+3\zeta_0,\label{4.28}\\
z'&&=z\left(4-x-y+3xw_0-3\zeta_0+w_1\kappa^2\right),\label{4.29}
\end{eqnarray}
where for the stationary points in the last equation above, we must
put $z_0=0$. We can see that the last equation is decoupled from the
system, and mainly governs the stability with respect to the
variable $z$ only. The stationary points of the system
(\ref{4.27})-(\ref{4.28}) are already found in section \ref{sec:2}
and these are presented in Eqs. (\ref{2.11.0.0})-(\ref{2.11.0.1}),
where we need to put $q=0$. The first fixed point should be excluded
from the future investigation, since it has $w_{eff}=\frac{1}{3}$.
Thus we have one possible fixed point candidate that may describe
inflation (let us denote it $P_{5a}$):
\begin{equation}
x_0=1-y_0,\,\,
y_0=\frac{1}{6w_0}\left [ 3w_0+w_1\kappa^2-3\zeta_0 \pm\sqrt{(3w_0+w_1\kappa^2-3\zeta_0)^2+36w_0\zeta_0} \right ],
\label{4.30}
\end{equation}
with,
 \begin{equation}
w_{eff}=\frac{1}{3}\left [ 3x_0w_0 +w_1\kappa^2 -3\zeta_0 \right ].
\label{4.31}
\end{equation}
Now in order to describe inflation, we need to construct a de Sitter
solution, which should be stable with respect to the coordinates
$x$, $y$ and unstable with respect to $z$, in order to provide an
exit from inflation. Thus let us put for (\ref{4.31})
$w_{eff}=-1+\gamma$, where $0<\gamma\ll 1$. In this case, the third
eigenvalue associated with the coordinate $z$ will exactly be
$\mu_3=\gamma$, so this provides an instability of de Sitter point.
Moreover deriving $x_0$ from (\ref{4.31}) and equating it to
(\ref{4.30}), we find an additional relation between the free
parameters,
 \begin{equation}
\zeta_0=\frac{(1-\gamma)(3-3\gamma+3w_0+w_1\kappa^2)}{3(1+w_0-\gamma)},
\label{4.31.1}
\end{equation}
or if we require that all the values of parameters are not very
small in comparison with $\gamma$, we have,
 \begin{equation}
\zeta_0=\frac{3+3w_0+w_1\kappa^2}{3(1+w_0)}-\frac{6+3w_0+w_1\kappa^2}{3(1+w_0)}\gamma.
\label{4.31.2}
\end{equation}

\subsubsection{Asymptotic State II: Present Time and Late-time Acceleration}

Now let us consider the present-time asymptotic case, in which case
instead of the dynamical system of Eqs. (\ref{4.24})-(\ref{4.26}) we
have the following system,
\begin{eqnarray}
x'&&=(x-1)(w_1\kappa^2+3xw_0-x)-x(y+3\zeta_0)-q,\label{4.32}\\
y'&&=-y^2+y(1-3\zeta_0+3xw_0-x+w_1\kappa^2)+3\zeta_0+q,\label{4.33}\\
z'&&=z\left(4-x-y+3xw_0-3\zeta_0+w_1\kappa^2\right),\label{4.34}
\end{eqnarray}
where for stationary points in the last equation we must require
$z_0\neq 0$. Thus, from the last equation we have
 \begin{equation}
y_0=4-3\zeta_0+w_1\kappa^2+x_0(3w_0-1).
\label{4.35}
\end{equation}
Substituting the above in Eq. (\ref{4.32}) we find,
$$
x_0=\frac{-q-w_1\kappa^2}{3(w_0+1)},
$$
and by substituting in Eq. (\ref{4.33}) we find,
$$
x_0=\frac{12\zeta_0+q-3w_1\kappa^2-12}{3(3w_0-1)}.
$$
In effect, the system will compatible only if the following
condition holds true,
$$
\frac{-q-w_1\kappa^2}{3(w_0+1)}=\frac{12\zeta_0+q-3w_1\kappa^2-12}{3(3w_0-1)},
$$
which gives us,
 \begin{equation}
q=\frac{3w_0+3-3\zeta_0+w_1\kappa^2-3\zeta_0w_0}{w_0},
\label{4.36}
\end{equation}
and the corresponding stationary point is (let us denote it
$P_{5b}$)
 \begin{equation}
x_0=\frac{3\zeta_0-w_1\kappa^2-3}{3w_0},\,\,y_0=\frac{3w_0+3-3\zeta_0+w_1\kappa^2}{3w_0},\,\,x_0+y_0=1.
\label{4.37}
\end{equation}
Combining (\ref{4.36}) with (\ref{4.31.2}) we find,
 \begin{equation}
q=(6+3w_0+w_1\kappa^2)\gamma,
\label{4.38}
\end{equation}
so we can see that $q$ is very small but does not
vanish.\footnote{Putting $w_{eff}=-1$ exactly for the first point,
we get $q=0$.} Note also that the effective EoS for this point,
calculated for (\ref{4.13.2}) yields exactly $w_{eff}=-1$ for any
values of the parameters.

\subsubsection{combining and stability analysis.}

By looking the differential equations in Eqs.
(\ref{4.27})-(\ref{4.28}) and (\ref{4.32})-(\ref{4.33}), it can be
seen that these are identical apart for some additive constants, and
in effect the eigenvalues will be identical for the two systems. Let
us denote $x'=f$ and $y'=g$ for notational simplicity. In effect,
the equation that determines the eigenvalues, takes the following
form,
\begin{equation}
\left |
\begin{array}{l}
(f_x)_0-\mu \hspace{1cm} (f_y)_0
\\
\\
\,\,\,\,\,\,\, (g_x)_0 \hspace{1cm}  (g_y)_0-\mu
\end{array}
\right | =0,
\label{4.39}
\end{equation}
where $(f_x)_0$, $(f_y)_0$, $(g_x)_0$ and $(g_y)_0$, are equal to,
$$
(f_x)_0=w_1\kappa^2+6w_0x_0-x_0-3\zeta_0-3w_0,
$$

$$
(f_y)_0=-x_0,
$$

$$
(g_x)_0=(3w_0-1)(1-x_0),
$$

$$
(g_y)_0=-1+x_0-3\zeta_0+3w_0x_0+w_1\kappa^2,
$$
and in the above relations we took into account that for both points
we have $x_0+y_0=1$.

The solution of Eq. (\ref{4.39}) for the fixed point $P_{5a}$ with
positive sign in (\ref{4.30}) reads,
 \begin{equation}
\mu_1=-\sqrt{(3w_0+w_1\kappa^2-3\zeta_0)^2+36w_0\zeta_0},\,\, \mu_2=\frac{1}{2}\left[ -2+ 3w_0+w_1\kappa^2-3\zeta_0 +\mu_1 \right],
\label{4.40}
\end{equation}
and for the point $P_{5b}$
 \begin{equation}
\mu_1=-4,\,\, \mu_2= -6 - 3w_0 - w_1\kappa^2+3\zeta_0,
\label{4.41}
\end{equation}
and by substituting the expression (\ref{4.31.2}) for $\zeta_0$, for
the fixed point $P_{5a}$, we have,
 \begin{equation}
\mu_1=-\frac{3 + 6w_0 + 3w_0^2 + w_0w_1\kappa^2}{w_0+1},\,\, \mu_2=-4,
\label{4.42}
\end{equation}
and for the point $P_{5b}$ we get,
 \begin{equation}
\mu_1=-4,\,\, \mu_2= -\frac{3 + 6w_0 + 3w_0^2 +
w_0w_1\kappa^2}{w_0+1}. \label{4.43}
\end{equation}
Note that we have set $\gamma=0$, since all the eigenvalues, even in
this case, do not vanish. Thus we can see that both points $P_{5a}$
and $P_{5b}$ are stable or unstable with respect to coordinates $x$
and $y$ simultaneously, and stability condition reads,
 \begin{equation}
(3 + 6w_0 + 3w_0^2 + w_0w_1\kappa^2)(w_0+1)>0.
\label{4.45}
\end{equation}
Moreover, we need to require that both fixed points lie in the
physical region, namely at $0\leqslant x_0\leqslant 1$ and
$0\leqslant y_0\leqslant 1$, so the following conditions must be
satisfied,
\begin{eqnarray}
w_1\kappa^2<-3,\,\,w_0\geqslant c,&&\label{4.46}\\
w_1\kappa^2=-3,\,\,w_0> 0,&&\label{4.47}\\
-3<w_1\kappa^2<0,\,\,c\leqslant w_0 <0 \| w_0>0,&&\label{4.48}\\
w_1\kappa^2>0,\,\,w_0\leqslant c,&&\label{4.49}
\end{eqnarray}
where,
$$
c=-\frac{1}{3}(w_1\kappa^2+3).
$$
In Table \ref{tab5} we have gathered all the fixed points and the
corresponding eigenvalues for the case of an oscillating form of the
interaction term $Q$.
\begin{table}[H]
\caption{\label{tab5}Fixed points for the case of oscillating $Q$.}
\begin{center}
\begin{tabular}{|c|c|c|c|c|c|c|c|}
\hline
$P_i$ & $x_0$ & $y_0$ & $z_0$ & $\mu_1$ &   $\mu_2$ & $w_{eff}$ \\
\hline
  5a  & $1-y_0$   & $\frac{1}{6w_0}\left [ 3w_0+w_1\kappa^2-3\zeta_0 \pm\sqrt{(3w_0+w_1\kappa^2-3\zeta_0)^2+36w_0\zeta_0} \right ]$   & $0$   & $-\frac{3 + 6w_0 + 3w_0^2 + w_0w_1\kappa^2}{w_0+1}$     & $-4$ & $-1+\gamma$  \\
\hline
  5b  & $1-y_0$   & $\frac{3w_0+3-3\zeta_0+w_1\kappa^2}{3w_0}$   & $\neq 0$   & $-4$  & $-\frac{3 + 6w_0 + 3w_0^2 + w_0w_1\kappa^2}{w_0+1}$ & $-1$  \\
\hline
\end{tabular}
\end{center}
\end{table}
It is worth discussing some interesting scenarios, so assume that
$w_0=1$, $w_1\kappa^2=-1$, then the fixed points and the eigenvalues
become,
$$
P_{5a}:\,\,\,x_0=\frac{1}{6},\,\,y_0=\frac{5}{6},\,\,\,\mu_1=-\frac{11}{2},\,\,\mu_2=-4,
$$

$$
P_{5b}:\,\,\,x_0=\frac{5}{6},\,\,y_0=\frac{1}{6},\,\,\,\mu_1=-4,\,\,\mu_2=-\frac{11}{2}.
$$
So by changing the values of the parameters $w_0$ and $w_1$, it is
possible to appropriately fix the fixed point $P_{5b}$, which recall
that it corresponds to late-time acceleration, in order some fixed
relation between $\rho_{dm}$ and $\rho_{de}$ holds true. Note that
in this section we studied only case for which the choices of the
functions $f(\rho)$ and $G(H^2)$ were the simplest choices, but in
principle a more involved functional form for these functions may
lead to more interesting phenomenology. In such a case though, the
analytical treatment will be possibly insufficient, so a concrete
numerical analysis will be needed.

In conclusion, we constructed a cosmological model that describes
the inflationary and the late-time acceleration era, due to an
oscillating interaction term between the dark energy and dark matter
fluids. We found two fixed points, with the first being $P_{5a}$,
which describes the initial de Sitter solution, namely the
inflationary era, and this fixed point was unstable, a feature which
indicates that the graceful exit is triggered. The second fixed
point was also found to be a de Sitter fixed point, which describes
the present day acceleration. As we demonstrated, in the oscillating
model we discussed, during the early Universe there was no
interaction between dark energy and dark matter, however during the
late-time the interaction was present. Finally, as we showed, by
appropriately choosing the parameters, it is possible to produce
some fixed relation for $\rho_{dm}/\rho_{de}$.

Before closing this section, let us briefly discuss an interesting
issue with regards the early-time era. It would be interesting to
calculate the slow-roll indices and the corresponding observational
indices for the inflationary era we presented in section \ref{infl}.
For example by assuming the perfect fluid approach
\cite{nojodineos1}, we can express the spectral index of primordial
curvature perturbations $n_s$ and the scalar-to-tensor ratio $r$ in
the usual way these are given in the case of a canonical scalar
field,
$$
n_s=1-6\epsilon+2\eta,
$$

$$
r=16\epsilon.
$$
with the slow-roll indices being defined in terms of the Hubble rate
as follows,
$$
\epsilon= -\frac{\dot H}{H^2},
$$

$$
\eta= \epsilon -\frac{\ddot H}{2H\dot H}.
$$
By combining Eqs. (\ref{4.1}) and (\ref{4.1.1}) it is easy to
calculate the slow-roll parameter $\epsilon$, which reads,
$$
\epsilon=\frac{1}{2}\left [ 4- y +x(3w_0-1) +w_1\kappa^2 -3\zeta_0 \right ].
$$
Differentiating (\ref{4.1.1}) and by combining with equations
(\ref{4.1})-(\ref{4.4}) we can find a similar expression for the
slow-roll parameter $\eta$. Note here that the resulting expression
for $\eta$ is much more complicated so we omit it. There is a major
obstacle in calculating the slow-roll indices however, since in the
expressions for $\epsilon$ and $\eta$, we must use values for $x$
and $y$ not on the stationary point, but from some point near the
stationary point. The resulting slow-roll indices must be expressed
in terms of the $e$-foldings number $N$, but doing this analytically
is a rather formidable task, that exceeds the purposes of this work.
A numerical approach though might be less difficult to perform, so
we hope to address this task in a future work.

\section{Conclusions} \label{sec:5}

In this work we analyzed in detail the phase space of a cosmological
system that contains cosmological fluids that have various forms of
equation of state. We firstly discussed the simplest forms of EoS,
and we found the fixed points of the cosmological dynamical system
and we discussed the physical significance of these fixed points. In
addition, we discussed theories that admit interactions between the
dark energy and dark matter fluids. In addition we introduced a new
class of interaction between dark energy and dark matter, in which
theories the interaction term is oscillating, allowing different
form of interactions for various eras during the cosmological
evolution. As we demonstrated, it is possible to have almost
negligible interactions at early times, that is, during the
inflationary era, and for the same model the interaction is turned
on at late times. The cosmological dynamical system of the
oscillating interaction term is in turn decomposed into two distinct
dynamical systems at early and late times. Interestingly enough the
two dynamical systems predict two de Sitter fixed points
corresponding to early and late times, with the early-time de Sitter
point being unstable in one of the coordinates, a feature which
indicates the possible exit from the inflationary era. This
framework gives us the interesting fluid-filled Universe evolution,
unifying the early-time acceleration with late-time acceleration.

The fluid description offers an alternative viewpoint in modern
cosmology, which may describe successfully many eras of our
Universe's evolution. The dynamical system approach offers many new
insights since the fixed points of the dynamical system reveal the
attractors of the whole theory and their stability indicates whether
these attractors are final attractors of the system. A compelling
extension of this work is to include Loop Quantum Cosmology effects
in the EoS, as was performed in Ref. \cite{oikonomouimperfect}, so
we defer this task to a future work. Another strong motivation to
adopt the fluid approach in cosmology, is the late-time acceleration
era, since this can be modeled by fluid cosmology
\cite{Balakin:2010xb}. Furthermore, as modified gravity maybe easily
presented in the effective fluid representation, the dynamical
systems approach turns out to be useful also for the study of the
cosmology in modified gravity. More interestingly, the dynamical
system approach can be applied to the study of bouncing cosmology,
described by a multi-component fluid. We aim to address this latter
issue in a future work. In conclusion, we believe that the
multi-component fluid cosmology may have some valuable insights to
offer to the cosmologists community.

\begin{acknowledgments}

This work was supported by the Russian Science Foundation (RSF)
grant 16-12-10401 (P.V.T.), by the Russian Foundation for Basic Research (RFBR) grant 17-02-01008 À (P.V.T.),
by MINECO (Spain), project
FIS2013-44881, FIS2016-76363-P and by CSIC I-LINK1019 Project
(S.D.O) and by the Russian Ministry of Education and Science,
Project No. 3.1386.2017 (S.D.O and V.K.O).

\end{acknowledgments}

\end{document}